\documentstyle[epsfig,subfigure,glas]{article}

\newlength{\dinwidth}                                                          
\newlength{\dinmargin}                                                         
\def\nostrocostrutto#1\over#2{\mathrel{\mathop{\kern 0pt \rlap 
  {\raise.2ex\hbox{$#1$}}}
  \lower.9ex\hbox{\kern-.190em $#2$}}}
\def\lsim{\nostrocostrutto < \over \sim}   

\newcommand{\be}{\begin{equation}}
\newcommand{\ee}{\end{equation}}
\newcommand{\ba}{\begin{eqnarray}}
\newcommand{\ea}{\end{eqnarray}}
\newcommand {\pom}  {I\hspace{-0.2em}P}
\newcommand {\alphapom} {\mbox{$\alpha_{_{\pom}}$}}

\catcode`@=11
\newcount\@tempcntc
\def\@citex[#1]#2{\if@filesw\immediate\write\@auxout{\string\citation{#2}}\fi
  \@tempcnta\z@\@tempcntb\m@ne\def\@citea{}\@cite{\@for\@citeb:=#2\do
    {\@ifundefined
       {b@\@citeb}{\@citeo\@tempcntb\m@ne\@citea\def\@citea{,}{\bf ?}\@warning
       {Citation `\@citeb' on page \thepage \space undefined}}%
    {\setbox\z@\hbox{\global\@tempcntc0\csname b@\@citeb\endcsname\relax}%
     \ifnum\@tempcntc=\z@ \@citeo\@tempcntb\m@ne
       \@citea\def\@citea{,}\hbox{\csname b@\@citeb\endcsname}%
     \else
      \advance\@tempcntb\@ne
      \ifnum\@tempcntb=\@tempcntc
      \else\advance\@tempcntb\m@ne\@citeo
      \@tempcnta\@tempcntc\@tempcntb\@tempcntc\fi\fi}}\@citeo}{#1}}
\def\@citeo{\ifnum\@tempcnta>\@tempcntb\else\@citea\def\@citea{,}%
  \ifnum\@tempcnta=\@tempcntb\the\@tempcnta\else
   {\advance\@tempcnta\@ne\ifnum\@tempcnta=\@tempcntb \else \def\@citea{--}\fi
    \advance\@tempcnta\m@ne\the\@tempcnta\@citea\the\@tempcntb}\fi\fi}
\catcode`@=12

\def\met{\mbox{${\hbox{$E$\kern-0.6em\lower-.1ex\hbox{/}}}_T$}} 
\def\D0{D\O}                            
\def\ppbar{$p\overline{p} $}            

\begin{document}
\begin{titlepage}{GLAS-PPE/97--02}{\today}
\title{QCD Effects in Hadronic Final States}
\centerline{ E. A. De Wolf (Universitaire Instelling Antwerpen), 
A. T. Doyle (University of Glasgow),}
\vspace{0.1cm}
\centerline{ N. Varelas (Michigan State University)
and D. Zeppenfeld (University of Wisconsin).}

\vspace{1cm}
\centerline{\it Summary of the Hadronic Final States Working Group}
\vspace{0.1cm}
\centerline{\it at the DIS97 Workshop, Chicago (April 1997).}

\vspace{1cm}
\begin{abstract}
Progress in the study of hadronic final states in deep inelastic scattering
as well as $p\bar{p}$, photoproduction and $e^+e^-$ annihilation,
as presented at the DIS97 workshop, is reviewed. 

\end{abstract}

\vspace{1cm}
\section*{Introduction}
The large centre of mass energies and increasing statistical precision 
available at HERA, the Tevatron and LEP combined with recent theoretical
developments open a new testing ground for QCD. The presentations on
hadronic final states presented at the DIS97 workshop place increasingly 
significant experimental and theoretical constraints on the strong interaction. 
This summary focuses on the common themes and highlights of the hadronic
final states working group.

\end{titlepage}

\section*{Fragmentation}
Perturbative QCD (pQCD) has proved to be a very successful theory in its 
application to hard processes. 
This enables the theory to be
employed as a tool to tackle more complicated
problems. One of these concerns the soft limit of QCD
and colour confinement.  Presently, multihadron production
phenomena cannot be derived in  a systematic way solely from
perturbation theory without additional model-dependent assumptions.

To investigate the limits of applicability of pQCD 
it is important to determine to what extent 
semi-soft phenomena in hard processes still reflect the properties of
the perturbative evolution phase. 
This line of research, initiated almost 15 years ago~\cite{dt,Dok92}, 
has reached a high level of sophistication 
(see~\cite{ochs9629} for  a recent review) 
and, as witnessed at this
workshop, continues to inspire analyses in all major high-energy experiments.
This summary deals with some of the most interesting results. 
 
\subsubsection*{Particle Rates}
In the current picture of hadron production, factorisation
plays a predominant role in the different evolution stages of the process:
`preparation' of the primary partonic configuration, additional
parton production described e.g. by (angular-ordered) parton  showers
(pQCD), hadron formation described e.g. by string or cluster
fragmentation (non-perturbative QCD), secondary hadronic cascade-decays
(QFD and non-perturbative QCD).
To test such an 
ansatz, a comparative
study of jet properties---including particle rates and spectra---in 
different reactions is required. 
At present this  has not yet become a topic of primary interest at
hadron-hadron colliders and  HERA in spite of its importance for QCD.

How much remains to be done is illustrated 
in~\cite{hemingway} where the
impressive results from the LEP experiments are updated and reviewed. 
As far as hadron production is concerned,  38 different inclusive production 
rates of mesons and baryons are now measured at the $Z^0$ 
and, for many of these, inclusive spectra are available.

In all, good agreement is observed for the rates with 
tuned versions of JETSET~7.4~\cite{jetset74} 
and HERWIG~5.9~\cite{herwig59}. 
A noteworthy
exception is the baryon sector which remains  an embarrassment for HERWIG.
Either a better retuning or a critical re-evaluation of the
cluster decay model seems required.
The Lund JETSET string approach fares better but contains a large number of
parameters related to flavour and spin. Since this number  is increasing
with time, little real predictive power is left.

In $e^+e^-$ annihilation at LEP, evidence for breaking of `jet universality'
and factorisation may have been found from 
excess $\eta$ production---above JETSET expectations---at large 
momentum in three-jet events~\cite{l3:eta} (glueball
production or surplus iso-singlet hadrons?) while no anomaly is seen for
$\pi^0$ production. 
It could be that the long-awaited direct
manifestation of gluon jet
fragmentation has finally been found~\cite{peterson:walsh}. If so,
even larger discrepancies could be expected for the $\eta'$~\cite{sjo_bruss}. 
The $f_0(975)$ and $a_0(980)$ mesons
could also play a special role in the dynamics of quark confinement
~\cite{fzero:azero}. A
comparative study of these and other hadrons in quark and in gluon jets
is called for.

Problems also appear  with strangeness production (mainly $K$ and $\Lambda$)
where DELPHI notes a deficit of strange particles in 
extreme two-jet events~\cite{delphi_9539}:
the production of strangeness depends on the {\em event
topology\/} in a manner that is not quantitatively described by JETSET.

Although HERA experiments have only started to investigate
the field so thoroughly explored at LEP, and information on
{\em identified\/} particles is still scarce, first
evidence has been found that the level of strangeness production in DIS
and photoproduction,
translated into a $s/u$ relative production rate is close to 0.2, to be
compared with 0.3 
in $e^+e^-$~\cite{h1zeus}.

Lessons to be learned from the vast amount of data in $e^+e^-$
annihilations at the $Z^0$ are that deviations from `universal fragmentation'
may well have been observed and that the topology of the
confining QCD fields is likely to play a role in hadroproduction.
The rich variety of such topological configurations possible
in $ep$ collisions poses a real challenge for the experimentalists.

\subsubsection*{Particle Spectra}
A striking prediction of the perturbative approach to QCD jet
physics is the depletion of soft particle production and the resulting
approximately Gaussian shape of the inclusive distribution in the variable
$\xi=\log E_{jet}/E$ for particles with energy $E$ in a jet of energy
$E_{jet}$---the famous ``hump-back plateau''~\cite{Dok92}.
Due to the {\em intrajet\/} coherence of gluon radiation, not 
the softest partons but those 
with intermediate energies ($E\propto E_{jet}^{0.3-0.4}$)
multiply most effectively in QCD cascades.


The shapes of  the measured particle energy spectra in $e^+e^-$ annihilation
turn out to be surprisingly close, over the whole momentum range  down to
momenta of a few hundred MeV, to the perturbative predictions based on the
Modified Leading Log Approximation (MLLA)~\cite{ochs9629}.
These observations can be taken as evidence that the perturbative phase
of the cascade development indeed leaves its imprint on the
final state hadrons. This, in turn, suggests that the
conversion of partons into hadrons occurs at a low
virtuality scale (of the order of the hadron masses), independent of the
scale of the primary process, and involves only  low-momentum transfers.
This  Local Parton-Hadron Duality (LPHD) may be connected
to pre-confinement properties of QCD which ensure that colour charges are
compensated locally~\cite{amati}. 
LPHD remains, however, a strong hypothesis
that is supposed to be valid only in an inclusive and average sense.
With LPHD, only two essential parameters are involved in the
perturbative description: 
the effective QCD scale $\Lambda$ and a (transverse momentum) 
cut-off parameter $Q_0$, resulting in a highly constrained theoretical
framework;  non-perturbative effects are essentially reduced to 
normalisation constants.

New data on charged particle spectra were presented at this workshop
by H1~\cite{deroeck}, ZEUS~\cite{bromley} and CDF~\cite{beretvas}. 
The HERA experiments concentrate
on the current
fragmentation region in DIS and perform the analysis in the
Breit frame, where the exchanged boson is completely spacelike.
The new data confirm with much increased statistical significance
the features observed in $e^+e^-$: approximately Gaussian shape of the
$\xi$ spectra with peak-position and width increasing with $Q$
as predicted in MLLA. Moreover, for sufficiently large $Q$,
they demonstrate the expected equivalence of the current region with
one hemisphere of an $e^+e^-$ event.
With increasing luminosity being accumulated at HERA, this work
should be extended to include moment and cumulant analyses
of the spectra for which detailed predictions exist~\cite{ochs9629}. 

\begin{figure}[htb]
  \centering
\mbox{
\subfigure[Evolution of $\xi$ with jet opening angle, $\Theta$,
           for $M_{\rm JJ}$~=~390~GeV.]
{\psfig{figure=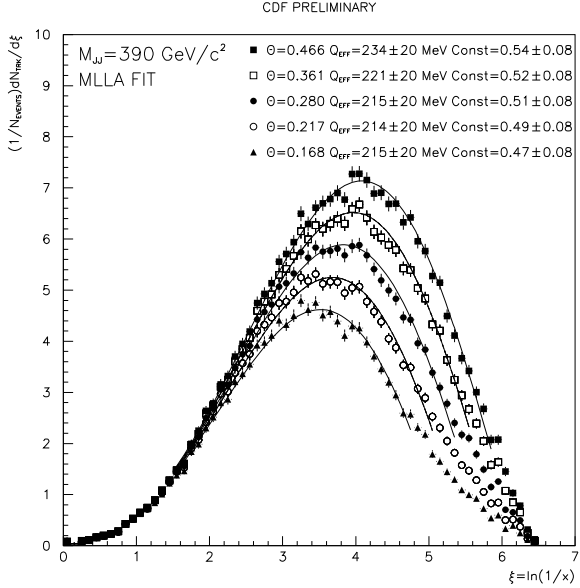,width=.45\textwidth}}\quad
        \subfigure[Evolution of the peak position with $M_{\rm JJ}\Theta$.]
{\psfig{figure=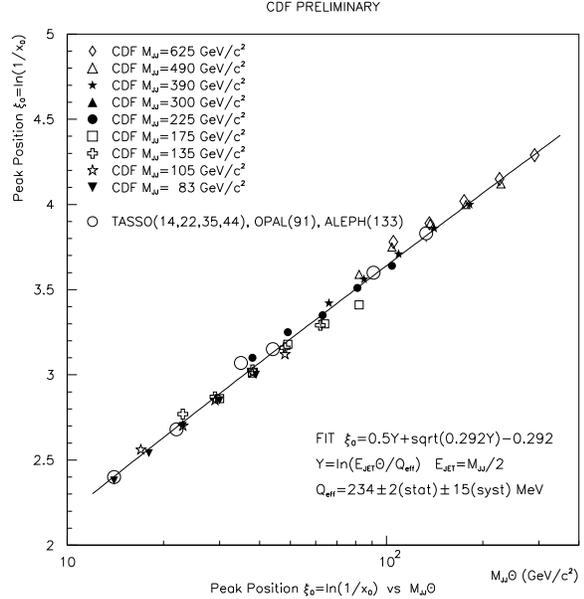,width=.45\textwidth}} 
}
  \caption[]{Comparison of preliminary CDF inclusive momentum distributions
             with MLLA predictions and $e^+e^-$ annihilation data.}
  \label{jet_frag}
\end{figure}

Beautiful confirmation of the  MLLA+LPHD approach has been presented 
by CDF~\cite{beretvas}. This experiment studies charged particle momentum
distributions in subsamples of dijet events. For fixed  dijet masses
(hence fixed jet energy) in the range $83<M_{\rm JJ}< 625$ GeV,
the $\xi$ distribution of tracks, within cones of various opening 
angle $\Theta$ (with respect to the jet axis), is 
studied (see Fig.~\ref{jet_frag}(a)).
As dijet mass $\times$ jet opening angle
increases, the peak of the spectrum, $\xi_o$,
shifts towards larger values
of $\xi$  in perfect agreement with MLLA predictions and
$e^+e^-$ data, as shown in Fig.~\ref{jet_frag}(b).
Similar analyses should be possible in DIS and photoproduction
at HERA but have not yet been attempted.

Although present data on charged particles appear to confirm strikingly 
the perturbative approach to soft hadronisation, the situation is
less clear-cut when spectra of identified particles/resonances are
examined. At LEP, the conclusion is unambiguous: the peak positions
do not agree with the naively expected mass dependence.
Here also, data on spectra of different hadron species and from different
jets at the Tevatron and HERA would be most helpful.
 
\subsubsection*{Limiting behaviour at low momenta}
\noindent The analytical perturbative approach allows one to predict the 
limit of
the  one-particle invariant  density in QCD jets  $E dn/{d^3p}\equiv
dn/dy\,d^2p_T$ at very small momenta $p$ or, equivalently, in the
limit of vanishing rapidity and transverse momentum~\cite{ochs9692}. 
If the dual description of hadronic and partonic states is 
adequate down to very small momenta, a finite, energy-independent limit of
the invariant hadronic density, $I_0$, is expected. 
This is a direct consequence of the colour coherence
in soft gluon branching. 
A possible rise  of $I_0$  with centre-of-mass energy
would indicate that either coherence or the local duality
(or both) break down. Since colour coherence is a general property of QCD as a
gauge theory, it is the LPHD concept that is tested in  measurements of the
soft hadrons.


The $e^+e^-$ annihilation data on charged and identified
particle inclusive spectra have been found to 
follow the MLLA  prediction surprisingly well, also at
low centre-of-mass energies. The invariant spectra at low momentum
scale approximately (within 10\%) between 1.6 and 140~GeV
and agree with  perturbative calculations  which become very sensitive
to the strong running of $\alpha_S$ at small scales~\cite{lo}. 

At this workshop, H1 presented the first Breit frame measurements
of the invariant energy spectra in DIS as a function of $Q$, 
(see Fig.~3 in~\cite{deroeck}).
For sufficiently high $Q$, the data  show that the low-momentum
limit in that region of phase space
is essentially independent of $Q$ and indeed similar 
to that in $e^+e^-$ annihilation.


\subsubsection*{Scaling violations}
Whereas the preceding paragraphs dealt mainly with the semi-soft limit of the
hadron spectra, promising results from H1 and ZEUS 
were presented to this workshop on the
scaling violations at larger values of  the scaled momentum
$x_p = 2p/Q$~\cite{deroeck,bromley}.
In QCD, scaling violations of the fragmentation function are expected, in full
analogy with scaling violations of structure functions,
due to increased gluon radiation. 
This leads to softer particle spectra with increasing energy. 
In principle, 
the scaling violations at large $x_p$ allow a
measurement of $\alpha_S$ and have been exploited for that purpose
in $e^+e^-$ annihilation. 
Whereas different $e^+e^-$ experiments
must be combined to cover a sufficient range in centre-of-mass energy, this
can be accomplished for $ep$ collisions in a  single experiment.

\begin{figure}[htb]
  \centerline{
   \psfig{figure=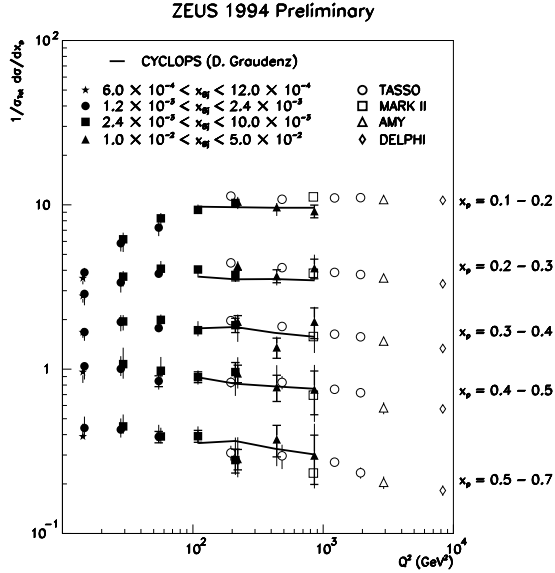,height=8cm,width=8cm}}
  \caption[]{$1/\sigma d\sigma/dx_p$ measurement as a function of $Q^2$.
             The preliminary ZEUS data are compared to data from $e^+e^-$
             annihilation experiments and the CYCLOPS NLO calculation
             of the inclusive charged hadron spectra.}
  \label{xp}
\end{figure}

H1 data provides evidence for violation of scaling in the current region of
the Breit frame. 
The corresponding ZEUS data, shown in Fig.~\ref{xp}, have also 
been compared with the CYCLOPS NLO calculation~\cite{CYCLOPS} incorporating 
fragmentation functions taken from $e^+e^-$~\cite{bromley}.
The data at large $x_p$  show  a weak dependence on the input proton
parton densities but a clear sensitivity to 
$\alpha_S$.
With analysis of more, already existing, data it should become possible to
use the complete NLO calculations to extract 
$\alpha_S$ in DIS.

\section*{Multiple Interactions}

At this workshop, 
CDF presented preliminary results on the measurement of events where two 
scattering processes occur in the same event~\cite{DPS}. 
For distinct processes $A$ and $B$, the cross section 
for this ``double parton (DP) scattering" is given by 
$\sigma_{\rm DP} = \sigma_A\sigma_B/\sigma_{\rm eff}$, where $\sigma_{\rm eff}$ 
is a process independent parameter.

Events were selected with a relatively low transverse energy ($E_T>$~16~GeV)
``photon" trigger 
in conjunction with three jets with $E_T>$~5~GeV.
The separation of DP events from the underlying QCD background is determined 
by studying variables sensitive to decorrelation effects.
In particular,
the azimuthal angle between the two best-balancing pairs (``photon"+jet versus
dijet) is approximately flat for the DP signal and enables a statistical 
separation of events.
A new feature of this analysis is that events with displaced
vertices, where the jets are reconstructed from separate origins,  
are used to evaluate $\sigma_A\sigma_B$
directly and hence reduce the theoretical uncertainties. 
This allows the first relatively precise determination of the effective 
cross section:
$$\sigma_{\rm eff}=(14.5\pm1.7^{+1.7}_{-2.3})~\rm{mb.}$$
No $x$-dependence is observed, within the uncertainty of $\simeq$ 20\%.
Assuming a uniformly
dense ball of partons and using the measured inelastic $p\bar{p}$ cross 
section, one expects $\sigma_{\rm eff}=$11~mb.
The measurement represents a milestone in the study of multiple interactions
and provides the first significant experimental constraint on such processes.

Multiple interactions, where two or more partons interact in the same event, 
represent a considerable uncertainty in the analysis of photoproduction
events at HERA. 
In particular, the extraction of the gluon content of
the photon at relatively low $x_\gamma$ requires careful modelling of 
these interactions, since they can contribute up to 50\% of the 
cross section at the relatively low $E_T$ values ($E_T>6$~GeV) measured
so far. 
The Tevatron result should aid in a realistic estimate of the 
uncertainties due to multiple interactions in the extraction of the 
gluon density of the photon at HERA. Similarly, such measurements 
improve background estimates to di-boson and boson+jet production at 
the Tevatron as well as the predictions of jet rates from multiple
interactions at the LHC.

\section*{Event Shapes}
The measurement of event shape
variables has been well-established in $e^+e^-$ annihilation
experiments. 
An important point in the development of our understanding of QCD is to ensure
that the measurement is well-defined theoretically at the required level of
precision. In this case 
variables are chosen which are relatively insensitive to soft gluon emission 
and collinear parton branching.
A determination of $\alpha_S(\mu)$ is therefore possible 
by comparison with NLO theory plus resummed series or NLLA calculations. 
At this workshop, impressive results from LEPII were presented which 
enabled a LEP average $\alpha_S(M_Z) = 0.120 \pm 0.005$ to be 
extracted~\cite{garcia}.

Recent theoretical developments in the understanding of infrared 
renormalon contributions, which lead to divergences from the 
perturbative calculations, allow the first steps to be made towards 
a direct comparison of theory and data without invoking hadronisation
models. These power corrections, with a characteristic 
$1/Q$ dependence, have been calculated for 
event shape variables~\cite{dasgupta} and could also be calculated for
differential jet rates.

\begin{figure}[htb]
  \centerline{
   \psfig{figure=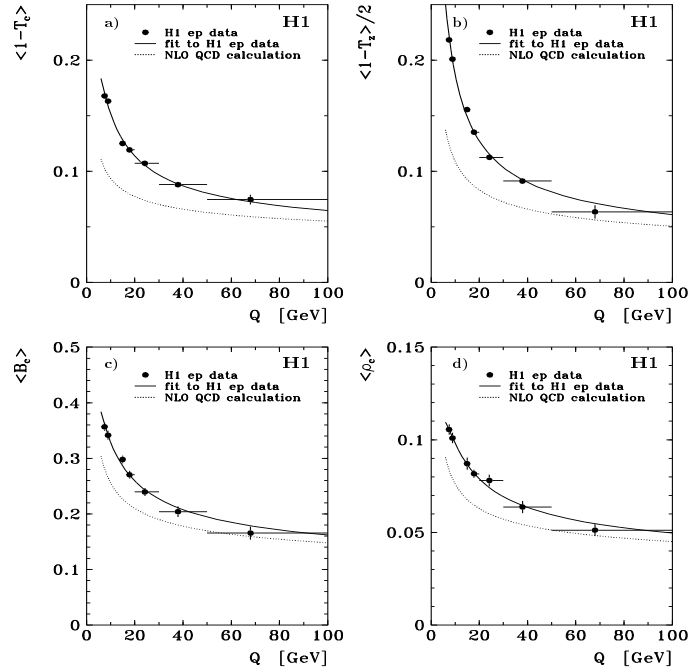,height=9cm,width=9cm}}
\vspace{0.5cm}
  \caption[]{Mean values of event shape variables as a function of Q,
             from H1. The values are for 1-thrust, calculated using (a) the thrust axis
             or (b) the photon axis, (c) the jet broadening and (d) the jet
             mass. 
             The dotted line indicates the NLO calculation.
             The full line indicates the fit incorporating power corrections.}
  \label{eshape}
\end{figure}

H1 presented new measurements of the thrust, jet broadening and
jet mass in DIS for momentum transfers, $7<Q<100$~GeV, in the 
current region of the Breit frame~\cite{rabbertz}. 
The mean values of the event shape data show similar trends to results
from $e^+e^-$ annihilation experiments as a function of $Q$.
In the DIS case, one advantage is that the event axis is determined by
the direction of the virtual boson, whereas in $e^+e^-$ annihilation
the axis has to be determined from the final state hadrons using 
e.g. the thrust axis.
The data, shown in Fig.~\ref{eshape},  
have been fitted to NLO theory plus the calculated power 
corrections of Dasgupta and Webber.
The important conclusion is that the size of the power correction,
characterised by the parameter $\bar{\alpha_o}$, 
is consistent with a single value of 
$\bar{\alpha_o} = 0.491\pm0.003$~(exp)~$^{+0.079}_{-0.042}$~(theory)
for three of the four event shape variables. In the case of jet broadening,
the calculation of the power corrections is subject to large uncertainties:
hence this particular variable does not satisfy the requirement of being 
theoretically well-defined and is not included in the global fit.
The development of these power corrections is not only intrinsically 
important, but should also enable more precise
extractions of $\alpha_S$ by constraining the 
hadronisation uncertainties more precisely.
\section*{Jet Shapes}
\begin{figure}[htb]
  \centerline{
   \psfig{figure=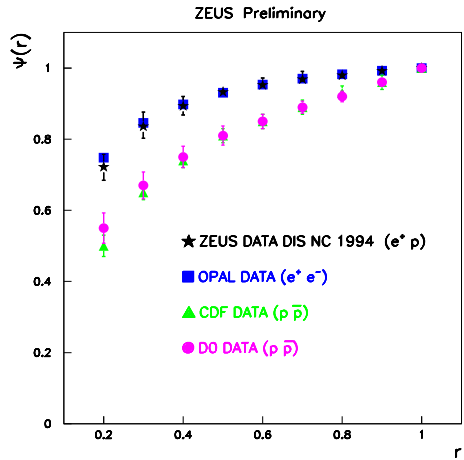,height=6cm,width=6cm}}
  \caption[]{Comparison of jet shape measurements from ZEUS(DIS), 
             OPAL($e^+e^-$), CDF and \D0 ($p\bar{p}$).
             The jet energy ranges are $37<E_T^{\rm jet}<45$~GeV,
            35~GeV$<E^{\rm jet}$, $40<E_T^{\rm jet}<60$~GeV and
            $45<E_T^{\rm jet}<70$~GeV, respectively.}
  \label{jshape}
\end{figure}

The shape of the transverse energy distribution of particles within
a jet produced in various interactions allows the primary parton 
source of the jet to be identified. In addition, the data provide strong 
constraints on the coherence properties of the showering partons and 
enable tests of the universality of the fragmentation process. 
In an analysis from the ZEUS Collaboration~\cite{SHAPES}, the jet shapes
measured in photoproduction and DIS were compared with those from $e^+e^-$
annihilation and $p\bar{p}$ experiments. Jets are measured using the cone
algorithm with a cone radius of 1. The jet shape, $\psi(r)$, is defined as
the average fraction of the jet's transverse energy that lies within an 
inner cone of radius $r$. The distributions shown in Fig.~\ref{jshape} are 
therefore integral plots
with $\psi(r)$=1 at r=1, whose rate of fall-off measures how broad the jet 
is. The data shown are for minimum jet transverse energies around 40~GeV.
It is observed that the DIS and $e^+e^-$ data contain $\simeq$ 70\% 
of their transverse energy within a sub-cone radius of 0.2, consistent with 
well-collimated quark jets. In contrast, the $p\bar{p}$ data jets are 
rather broad, with only $\simeq$ 50\% of their transverse energy being 
contained within the same sub-cone radius, consistent with predominantly
gluon jets in this $E_T$ range.

Photoproduction data (see Fig.~2 in~\cite{SHAPES}) were also studied 
as a function of 
pseudorapidity and transverse energy. The observed changes in jet shape
were reproduced in models which incorporate both direct and resolved photon
processes provided that the resolved processes include the multiple 
interactions discussed above. 
NLO calculations from Klasen and Kramer~\cite{klasen} 
determine the jet shape only at the lowest non-trivial order.
In order to describe the data,
an $R_{\rm sep}$ parameter is 
introduced which determines when two partons are merged into a single jet.
The jet shape distribution is well described by NLO calculations with an 
$R_{\rm sep}$ parameter which increases with increasing rapidity in the proton
direction, but which is in the
range $1.3 < R_{\rm sep} < 1.8$.
Differential distributions of the average transverse energy in intervals
of cone radius will enable $R_{\rm sep}$ to be fitted and provide further 
constraints on the models.

\section*{High-$E_T$ Jet Results from the Tevatron}

The Tevatron
Collider provides a unique opportunity to study the properties of hard
interactions in \ppbar\ collisions at short distances.  The production of jets
at large $E_T$ and its comparison with perturbative QCD calculations are of
interest as they can serve as a test of the elementarity of the
partons.  
    
    The CDF and D\O\ collaborations have measured jet cross sections over ten
orders of magnitude in $d^2\sigma/dE_Td\eta$ up to $E_T = 500$ GeV, half way to 
the kinematic limit.  The challenge of measuring such a steeply 
falling spectrum is the understanding of the energy calibration of
jets.  The highest $E_T$ jets are not directly calibrated, resulting
in large uncertainties.  In this kinematic region
the NLO calculations are well understood 
at the 10--20\% level.  However, precise
knowledge of the parton distribution functions in the proton 
is required before firm conclusions can be drawn from the comparison of data
and theory.
Collider data can 
constrain the parton distribution functions in the
proton and especially the gluon distribution at moderate $x$.  Kosower 
presented a formalism to make such an extraction possible using NLO 
calculations, while minimising the amount of numerical computation 
involved~\cite{Kosower}.

\begin{figure}[htb]
  \centering
\mbox{
\subfigure[CDF data vs. theory]
{\psfig{figure=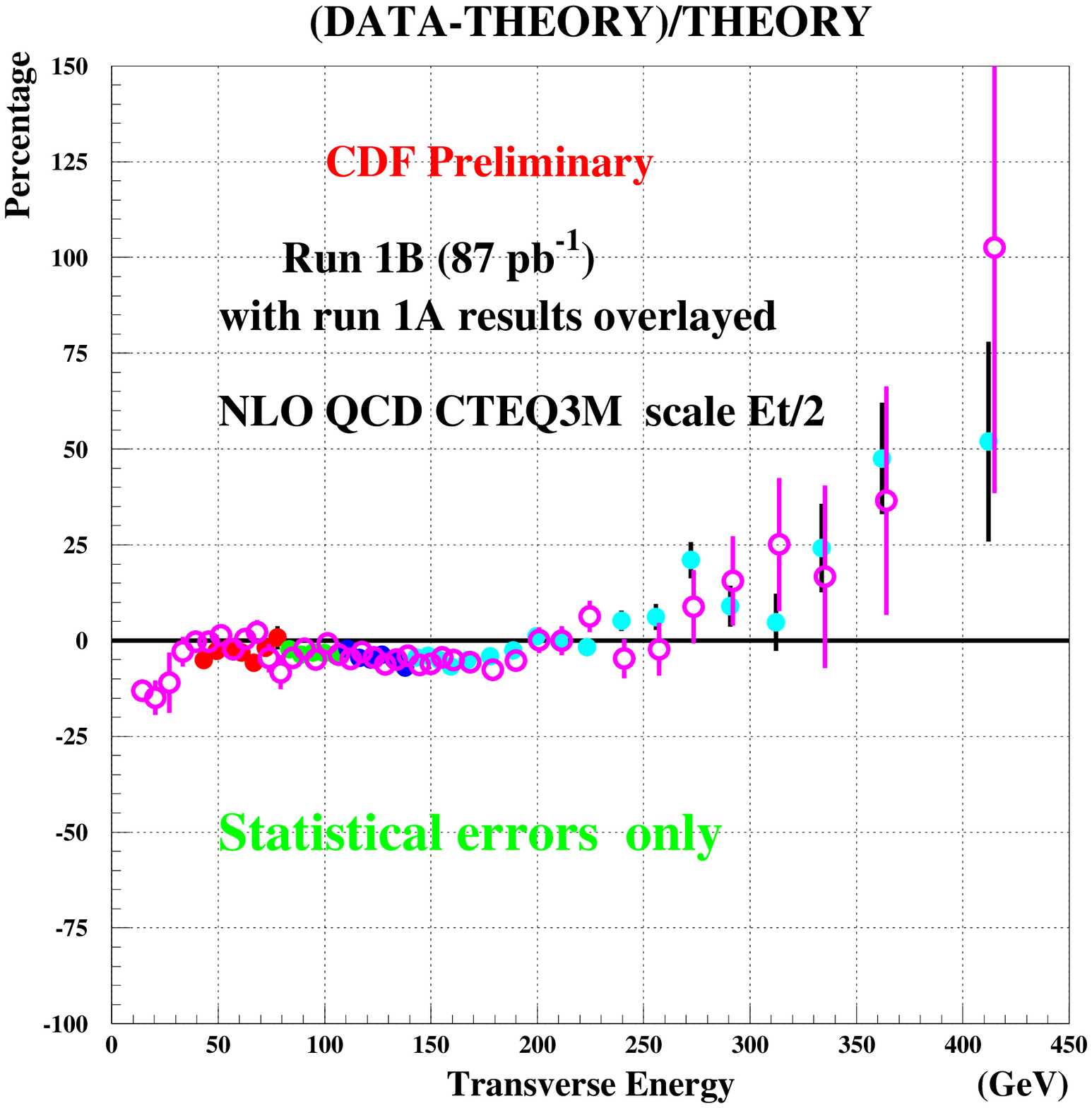,width=.45\textwidth}}\quad
        \subfigure[D\O\ data vs. theory ]
{\psfig{figure=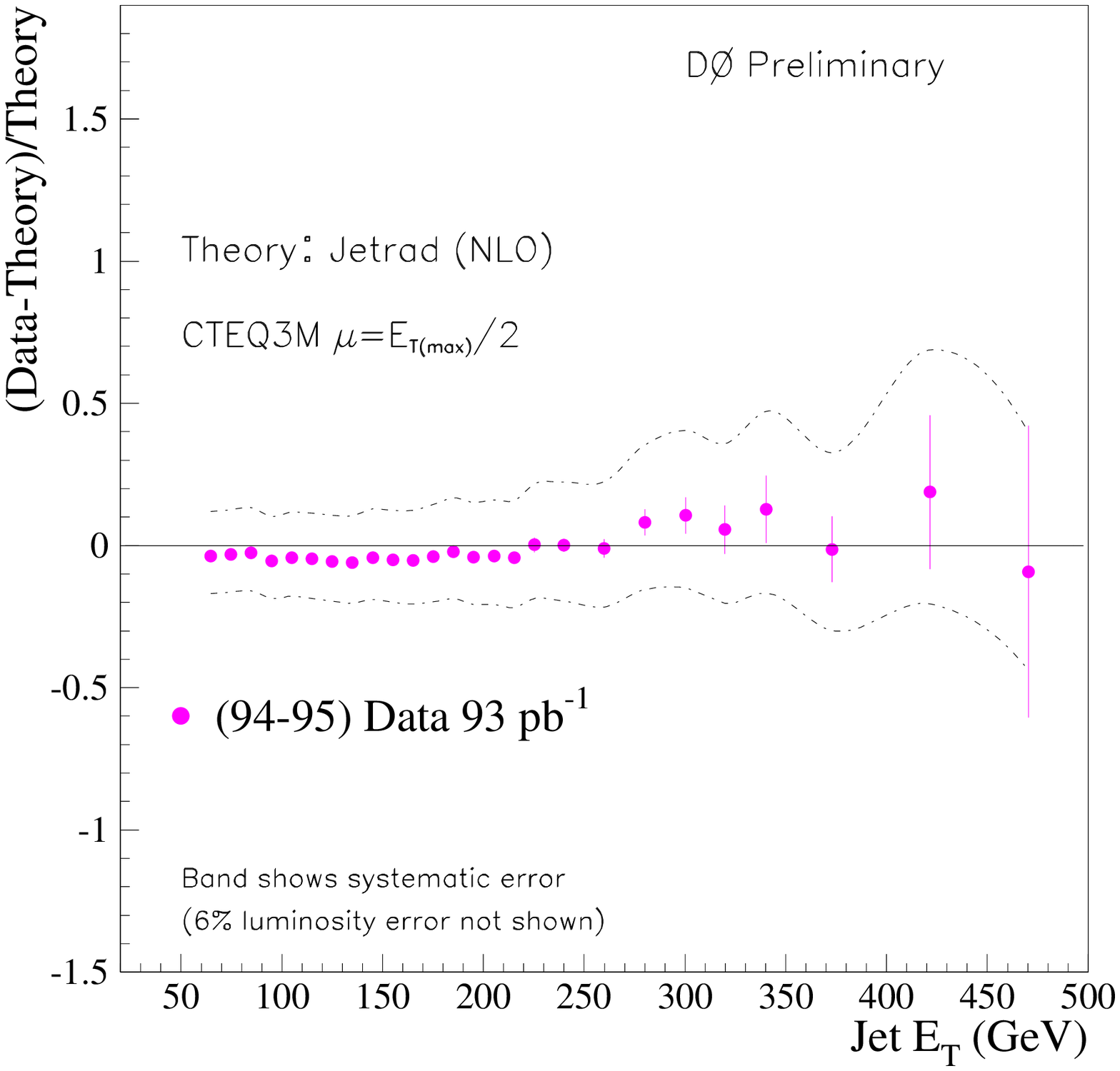,width=.45\textwidth}} 
}
  \caption[]{Ratio between experiment and theory for the inclusive jet cross
  section as measured by CDF and D\O.}
  \label{high_et_jets}
\end{figure}

    The preliminary (published) inclusive jet cross sections as measured by 
CDF using the 1994--1995 (1992--1993) data sample in the pseudorapidity
region of $0.1\leq|\eta|\leq0.7$ are compared to the NLO QCD in 
Fig. \ref{high_et_jets}(a) \cite{Hirosky,Chlebana}.  
The latter are based on calculations by 
EKS~\cite{eks} 
with {\small {CTEQ3M}}~\cite{cteq3m} parton densities, renormalisation
and factorisation scales $\mu=E_T^{\rm  jet}/2$, and the standard Snowmass jet
cone algorithm.
The data and the prediction are in excellent 
agreement for $E_T < 250$~GeV; at higher $E_T$, however, the data lie
significantly above the predictions.  

    D\O\ presented updated inclusive jet cross sections
in the region of $|\eta|\leq0.5$ with significantly 
reduced systematic uncertainties (by about a factor of two), 
coming from a re-evaluation of the jet energy
scale corrections~\cite{Hirosky}.  
As shown in Fig.~\ref{high_et_jets}(b), these results are in
excellent agreement with NLO QCD over the entire $E_T$ range.  D\O\ compares 
the data to NLO QCD predictions using {\small {JETRAD}}~\cite{jetrad}, 
the {\small {CTEQ3M}} parton densities, the 
renormalisation and factorisation scales 
$\mu=E_T^{\rm  max}/2$, and a modified Snowmass jet cone algorithm with 
$R_{\rm sep}$=1.3.

    It should be noted that the D\O\ and CDF data have been compared to NLO 
QCD with slightly different input parameters which can introduce 
an $E_T$-dependent variation of $\simeq~10\%$ on the theoretical 
predictions~\cite{Hirosky}.
Also the two measurements probe different $\eta$ regions.

\begin{figure}[htb]
  \centerline{
   \psfig{figure=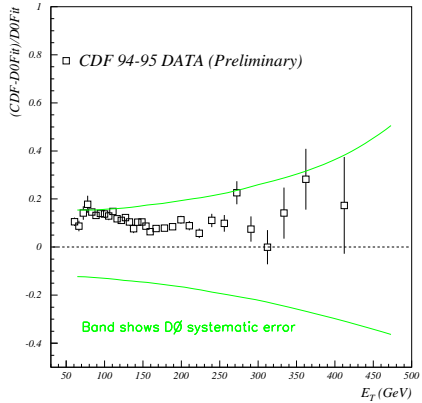,height=6cm,width=6cm}}
  \caption[]{Residual plot of the CDF data with a fit on the 
             $0.1\leq|\eta|\leq0.7$ D\O\ data.  The band shown represents the 
             D\O\ systematic uncertainty.}
  \label{cdf_d0_comp}
\end{figure}

In order to 
compare directly the results of the two experiments,
D\O\ also performed the analysis in the CDF $\eta$ region.
Figure \ref{cdf_d0_comp} shows the CDF data points as compared to a fit of the 
D\O\ data in the 
$0.1\leq|\eta|\leq0.7$ region.  The error band corresponds to the D\O\
systematic error which is mainly due to the jet energy scale uncertainty.  The
CDF data lie above the D\O\ fit but are within the experimental uncertainties.  
For a more 
quantitative comparison between the two experiments,
the correlations in the systematic uncertainties of the two data sets
must be taken into account.

\begin{figure}[htb]
  \centerline{
   \psfig{figure=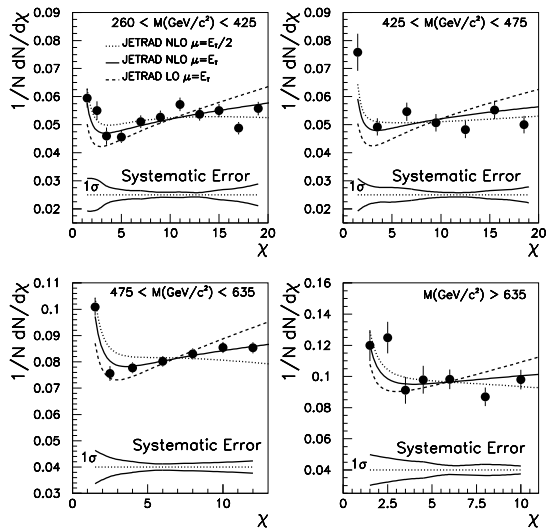,height=8cm,width=8cm}}
  \caption[]{Dijet angular distributions for D\O\ data compared to 
             {\small {JETRAD}} for LO and NLO predictions with two
             different renormalisation/factorisation scales.}
  \label{d0_chi}
\end{figure}

    The dijet angular distribution is an ideal tool to determine whether any
possible excess of events in high--$E_T$ inclusive jet production is 
due to new physics effects.
The angular distribution of the outgoing
partons is strictly governed by the helicities of the partons
participating in the hard process
and is relatively insensitive to the parton densities.  
Any unusual contact interaction (with effective scale $\Lambda$) will 
flatten the centre of mass 
scattering angle distribution (or create an excess of events at low $\chi$).  
The CDF published results on dijet angular distributions give a lower limit
of $\Lambda > 1.8$ TeV.  
Figure \ref{d0_chi} shows the
recent D\O\ $\chi$ distributions which are in good agreement 
with NLO QCD~\cite{Hirosky}.  
Using these data, 
D\O\ rules out at 95\% CL a model where quarks couple with a universal
contact interaction of scale $\Lambda\sim 2.1$~TeV.

\section*{Dijet Production in DIS at HERA}
A major area of study at HERA is dijet production in DIS,
which has earlier been used to determine $\alpha_S$ and constrain the
gluon density.
Previously, data have been compared to NLO semi-analytic calculations
using JADE-type algorithms. Now, with the flexible NLO Monte Carlo
programs MEPJET~\cite{mepjet} and DISENT~\cite{disent} available,
comparisons with various jet schemes are possible.
Results were presented at this workshop using the cone~\cite{mikunas,wobisch},
JADE~\cite{weber} and $k_T$~\cite{zomer} algorithms.

In order to measure the cross sections, detailed comparisons with models
incorporating parton showers/dipole chains and a hadronisation phase
have been made.
In general, these data are well described by the ARIADNE~\cite{ariadne} 
program and are reasonably 
well described by LEPTO~\cite{lepto} or HERWIG~\cite{herwig59}. 
The next stage in the development of ARIADNE
by the Lund group is the Linked Dipole Chain (LDC) model, which was 
reported at this workshop~\cite{hamid}.

A problem highlighted at the workshop relates to various attempts which 
have been
made to correct to $parton$ level in an attempt to determine the gluon
density or the strong coupling constant directly. However, the relationship
between the NLO partons and ARIADNE/LEPTO/HERWIG partons is far from clear and
this introduces an uncertainty for theorists who wish to
compare with published data. A presentation of the data corrected to hadron
level is therefore required. 

The general observation in various analyses,
with a range of different kinematic cuts,
is that the measured dijet cross sections/rates tend 
to be higher than those predicted by the NLO calculations incorporating a 
default coupling constant and parton densities which describe the total 
DIS cross sections\footnote{ 
A similar excess is observed in the ZEUS ``resolved" photoproduction 
dijet data when compared to the NLO calculations for the lowest 
$E_T>6$~GeV and relatively low $x_\gamma$ ($0.3<x_\gamma^{OBS}<0.7$) data
(see Fig. 2 in~\cite{saunders}).}. 

\begin{figure}[htb]
  \centerline{
   \psfig{figure=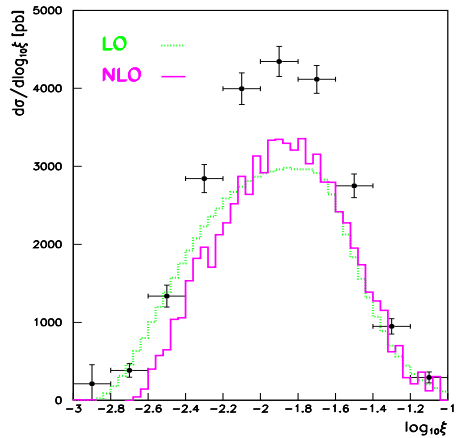,height=6cm,width=6cm}}
  \caption[]{ZEUS preliminary dijet cross section as a function of $\xi$,
             compared to LO and NLO predictions.}
  \label{ZEUS_xi}
\end{figure}

This is illustrated in Fig.~\ref{ZEUS_xi} where the ZEUS preliminary
dijet cross 
section, corrected to parton level, is compared to the NLO and LO 
predictions~\cite{mikunas}.
The cross section is measured as a function of $\xi = x(1+M_{\rm JJ}/Q^2)$, 
the momentum fraction of the parton entering the hard scattering process.
The data are $\simeq$ 30\% higher than 
the NLO calculation and this difference persists after taking into account
variations in calorimeter energy scale, jet energy resolution, the Monte 
Carlo used to correct to parton level, the input parton densities or the 
factorisation/renormalisation scale.
However, the shape of the cross section is well described by the NLO
calculations and this can be used to extract the power dependence 
of the gluon at
low-$\xi$, $\xi g(\xi) \propto \xi^{-\lambda}$. This results in a value of 
$\lambda = 0.38 \pm 0.04 \pm 0.18$ at $Q^2=4$~GeV$^2$.

\begin{figure}[htb]
  \centering
\mbox{
\subfigure[$h(x)$ power correction term.]
{\psfig{figure=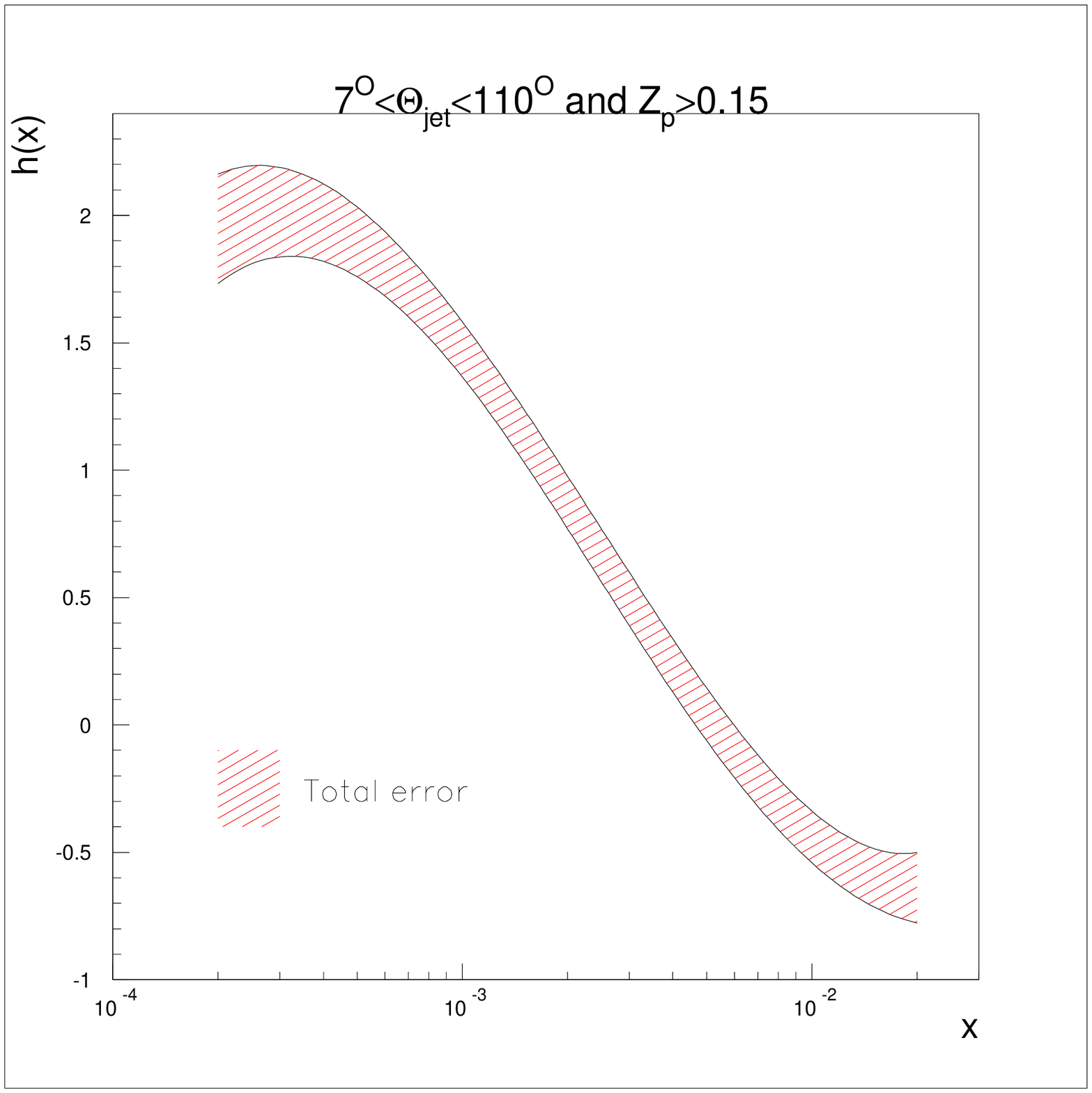,width=.45\textwidth}}\quad
        \subfigure[dijet rates compared to the global fit
                   incorporating the $h(x)$ power correction.]
{\psfig{figure=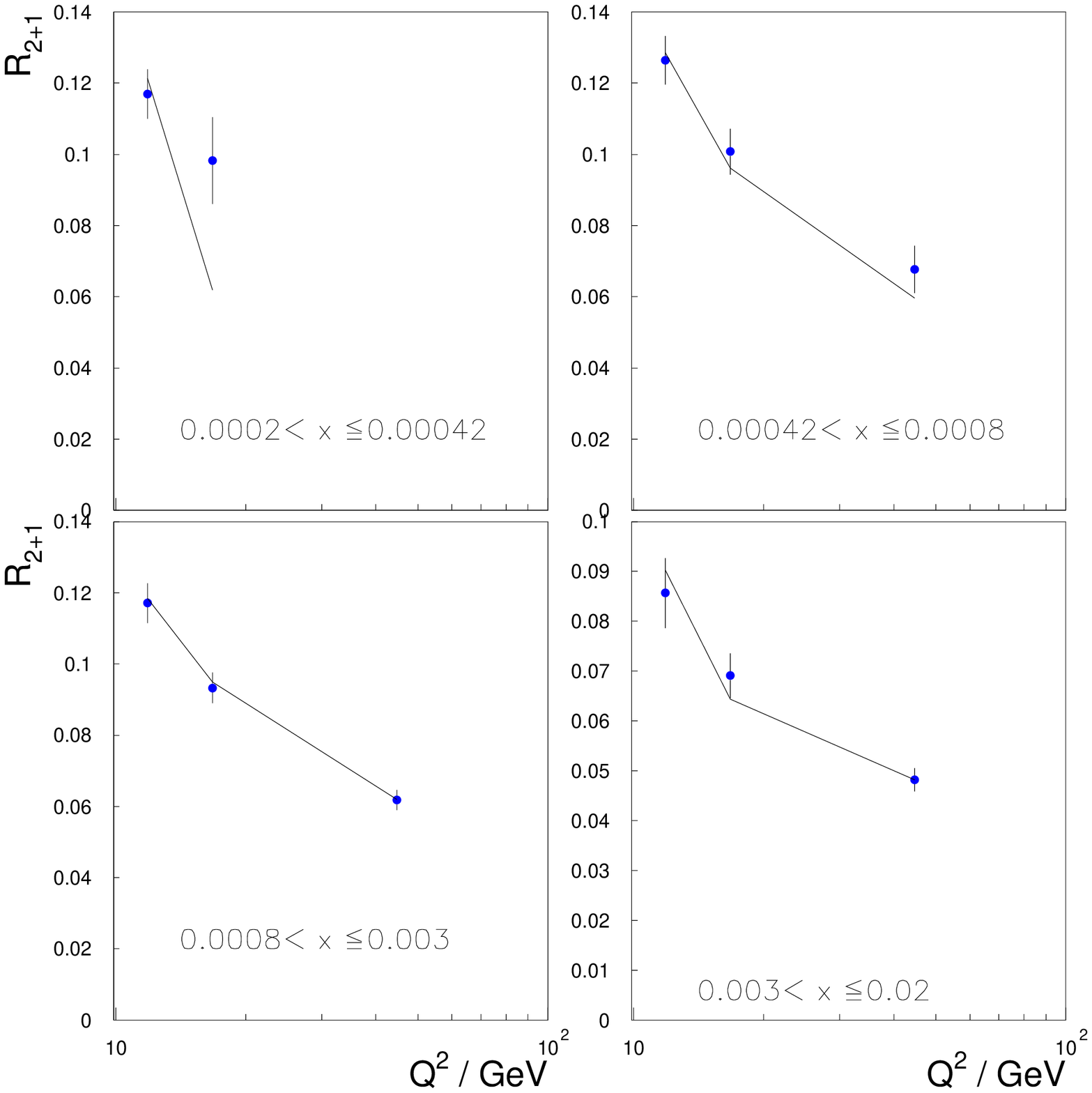,width=.45\textwidth}} 
}
  \caption{Analysis of H1 dijet data.}
  \label{jetfit}
\end{figure}

A further development has been taken in the H1 analysis. Using the $k_T$
algorithm in the Breit frame~\cite{zomer}, a global fit has been 
performed of the H1 and NMC DIS cross section measurements as well as 
the H1 preliminary dijet rates. 
In order to account for hadronisation effects, an additional power 
correction term is incorporated into the fit of the dijet rates. 
Although the functional form of these power corrections has not 
yet been calculated, it is clear from the fits to the data that such 
a term is required.
An empirical function $h(x) = \alpha+\beta\ln(x/x_o)
+\gamma\ln^2(x/x_o)+\delta\ln^3(x/x_o)$ is introduced, 
where $\alpha, \beta, \gamma$ and  $\delta$ are additional parameters in 
the fit and $x_o=10^{-4}$.
The additional contribution to the dijet cross section is determined as
$\Delta\sigma(x,Q^2) = h(x)/Q^2$. The fitted form of $h(x)$ is 
shown in Fig.~\ref{jetfit} together with the results of the global
fit incorporating this power correction term.
A calculation of the power corrections would therefore enable a simultaneous 
determination of its magnitude and provide further constraints on
$\alpha_S$ as well as the parton densities. 

\section*{BFKL-motivated measurements}
\setlength{\unitlength}{0.7mm}
\begin{figure}[htb]
\epsfxsize=2.5in
\epsfysize=2.5in
\begin{center}
\hspace*{0in}
\epsffile{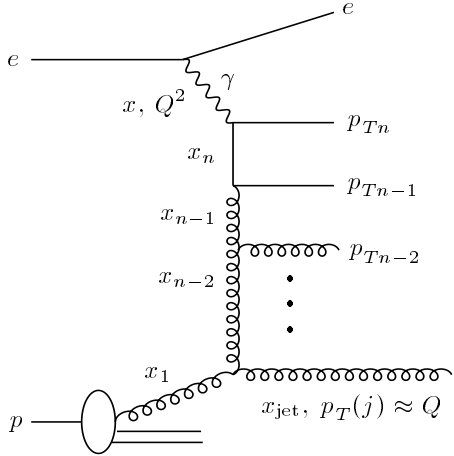}
\vspace{0.5cm}
\caption{
Gluon ladder diagram contributing to jet production in DIS. The position
and kinematics of the parton which can give rise to the forward jet is 
indicated. 
\label{fig:feyn}
}
\end{center}
\end{figure}

Recently, much interest has been focused on the small
Bjorken-$x$ region, where one would like to distinguish 
BFKL~\cite{bfkl} 
from the more traditional DGLAP evolution 
equation~\cite{dglap}. 
One of the dominant Feynman graphs responsible 
for parton evolution in DIS
is shown in Fig.~\ref{fig:feyn}. The $x_i$ denote the momentum fractions 
(relative to the incoming proton) of the incident virtual partons 
and $p_{Ti}$ is the transverse momentum of emitted parton $i$. 
Such ``ladder-type'' diagrams with strong ordering in transverse 
momenta, $Q^2\simeq p_{Tn}^2 \gg \ldots \gg p_T(j)^2$, but only soft 
ordering for the longitudinal fraction $x_1 > x_2>\ldots> x_n\simeq x$ 
are the source of the leading log $Q^2$ contributions which are summed in the 
DGLAP evolution equation~\cite{dglap}. 
In the BFKL approximation, transverse momenta are no longer 
ordered along the ladder while there is a strong ordering in the fractional 
momentum $x_n \ll x_{n-1}\ll\ldots \ll x_1\simeq x_{jet}$. 

BFKL evolution can be enhanced and DGLAP evolution suppressed by studying 
DIS events which contain an identified jet of longitudinal momentum 
fraction $x_{jet}=p_z(j)/E_{proton}$ (in the proton direction) 
which is large compared to Bjorken $x$~\cite{mueller}.
Furthermore, tagging a forward jet 
with $p_T(j)\simeq Q$ allows little room for
DGLAP evolution while the condition $x_{jet}\gg x$ leaves BFKL evolution 
active. Assuming BFKL dynamics leads to an enhancement of the forward jet 
production cross section proportional to $(x_{jet}/x)^{\alphapom -1}$,
where $\alphapom$ is the BFKL pomeron intercept, compared to 
the ${\cal O}(\alpha_S^2)$ QCD calculation with DGLAP evolution~\cite{MZ-prl}.

\begin{figure}[t]
\epsfxsize=5.0in
\epsfysize=3.0in
\begin{center}
\hspace*{0in}
\epsffile{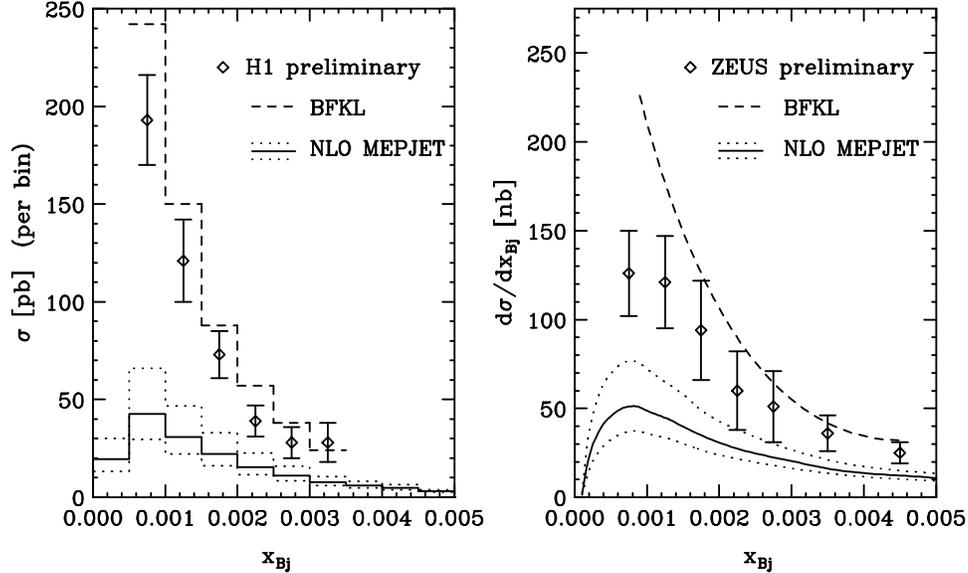}
\vspace*{0.5cm}
\caption{
Forward jet cross section at HERA as a function of Bjorken $x$ within (a) 
the H1~\protect\cite{H1-forward} and (b) the ZEUS~\protect\cite{woelfle}
acceptance cuts.
The BFKL result of Bartels et al.~\protect\cite{bartelsH1} 
is shown as 
the dashed line.
The solid  and dotted lines give the NLO MEPJET result for the scale 
choice $\mu_R^2=\mu_F^2=\xi(0.5\sum k_T)^2$ with $\xi=0.1, 1$ and 10, 
which provides a measure for the uncertainty of the NLO prediction. 
\label{fig:h1comp}
}
\vspace*{-0.1in}
\end{center}
\end{figure}

In Fig.~\ref{fig:h1comp}, recent data from H1~\cite{H1-forward} and 
ZEUS~\cite{woelfle} are compared with BFKL predictions~\cite{bartelsH1} 
and fixed order QCD predictions as calculated with the MEPJET~\cite{mepjet}
program at NLO. The conditions $p_T(j)\simeq Q$ and 
$x_{jet}\gg x$ are satisfied in the two experiments by slightly different
selection cuts. H1 selects events with a forward jet of $p_T(j)>3.5$~GeV 
(in the angular region $7^o < \theta(j) < 20^o$) with
\begin{equation}
   0.5  <  p_T(j)^2/Q^2\; < \; 2\;, \qquad \qquad
   x_{jet}  \simeq  E_{jet}/E_{proton} > 0.035\;; \label{eq:fj-H1}
\end{equation}
while ZEUS triggers on somewhat harder jets of $p_T(j)>5$~GeV 
and $\eta(j)<2.4$ with
\begin{equation}
   0.5  <  p_T(j)^2/Q^2\; < \; 4\;, \qquad \qquad
   x_{jet}  =  p_z(j)/E_{proton} > 0.035\;. \label{eq:fj-ZEUS}
\end{equation}

Clearly, both experiments observe substantially more forward jet events 
than expected from NLO QCD. A very rough estimate of the uncertainty of 
the NLO calculation is 
provided by the two dotted lines, which correspond to variations 
by a factor 10 of the renormalisation and factorisation scales
$\mu_R^2$ and $\mu_F^2$. 
A recent BFKL calculation (dashed lines) agrees better with the data, but 
here the overall normalisation is uncertain and the agreement may be 
fortuitous. Also, we recall that both experiments observe more centrally 
produced dijet events than predicted by the NLO QCD calculations. Whatever
mechanism is responsible for the enhancement in central jet production 
may also play a role in the enhanced forward jet cross section.
Clearly these issues must be resolved before the evidence for BFKL dynamics
can be elevated to the status of discovery.

The multiple gluon emission in ladder-type diagrams 
is also studied in jet-jet decorrelations at the Tevatron.
D0 presented preliminary results as a function of the 
pseudorapidity separation of the two leading jets in an event~\cite{BFKLD0}.  
The measurement is compared
to HERWIG and PYTHIA~\cite{jetset74} parton-shower Monte Carlo simulations, and to
BFKL predictions. The soft gluon
emissions are expected to decorrelate the transverse energy ($E_T$) and
azimuthal angle ($\phi$) of the produced jets as the rapidity interval
between them increases.
{\small {HERWIG}} and {\small {PYTHIA}} simulations reproduce the observed
decorrelation reasonably well.  However, the leading-log {\small {BFKL}} 
resummation~\cite{delduca}
predicts a larger decorrelation while a NLO QCD calculation underestimates
the decorrelation effects.  
Therefore, no clear conclusion on the question of BFKL dynamics can be 
drawn from the present Tevatron data.


\newpage
\section*{Instantons at HERA}
Perturbative QCD successfully describes hard scattering processes.
Beyond these, nonperturbative processes are predicted by QCD as well,
e.g. processes mediated via instanton configurations
in the path integral~\cite{instantons}. Of particular interest in $ep$
collisions are instanton processes which simultaneously produce $n_f$
light $\overline{q_L}q_R$ pairs and hence violate chirality by 
$\Delta Q_5=2n_f$ units. Ordinarily these processes are exponentially
suppressed, by a factor ${\rm exp}[-4\pi/\alpha_S]$. In conjunction 
with multiple gluon emission, however, this suppression is ameliorated
by a factor ${\rm exp}[-4\pi F(x')/\alpha_S(\mu)]$, where the so-called
``holy-grail function'' $F(x')$ equals unity at $x'=1$. $F(x')$ is known to 
decrease with decreasing $x'$, to about 1/2 at $x'\simeq 0.2$ but is not 
reliably calculable for small values of $x'$~\cite{instantons}. 

The expected fraction, $f^{(I)}$, of instanton 
induced events, compared to generic DIS
events at the same $x$ and $Q^2$, depends critically on the shape of the 
holy-grail function at small $x'$. Expectations range between 
$10^{-6}<f^{(I)}<10^{-3}$ if $F(x')$ approaches a constant below 
$x'_{min}=0.2\dots 0.3$~\cite{instantons}. Because several $q\bar q$ pairs
and gluons are produced isotropically, the striking signature of 
instanton-induced events would be very high particle 
multiplicity and high average
transverse energy deposition over a large region of the 
available phase space.

H1 reported on a search for such events~\cite{carli}.
The best limits are obtained from the non-observation of events with large
charged particle multiplicities. For 80~$<W<$~220~GeV, limits of 
$f^{(I)}\lsim 0.5\%$ have been set~\cite{carli}. 
These limits are still 
about one order of magnitude larger than expectations from instanton 
calculations~\cite{instantons}, but they begin to probe the interesting 
parameter range.
\newpage
\section*{$W + $ Jets Production}

    Hadronic production of $W$ and $Z$ bosons provides a clean probe 
of perturbative QCD calculations.  Three analyses were
presented from the D\O\ and CDF collaborations utilising the large 
sample of \ppbar\ collisions accumulated from the 1994-1995 Tevatron Collider 
run~\cite{Minor}.

\begin{figure}[htb]
  \centering
\mbox{
\subfigure[Cross sections for $W/Z + \geq n$ Jets]
{\psfig{figure=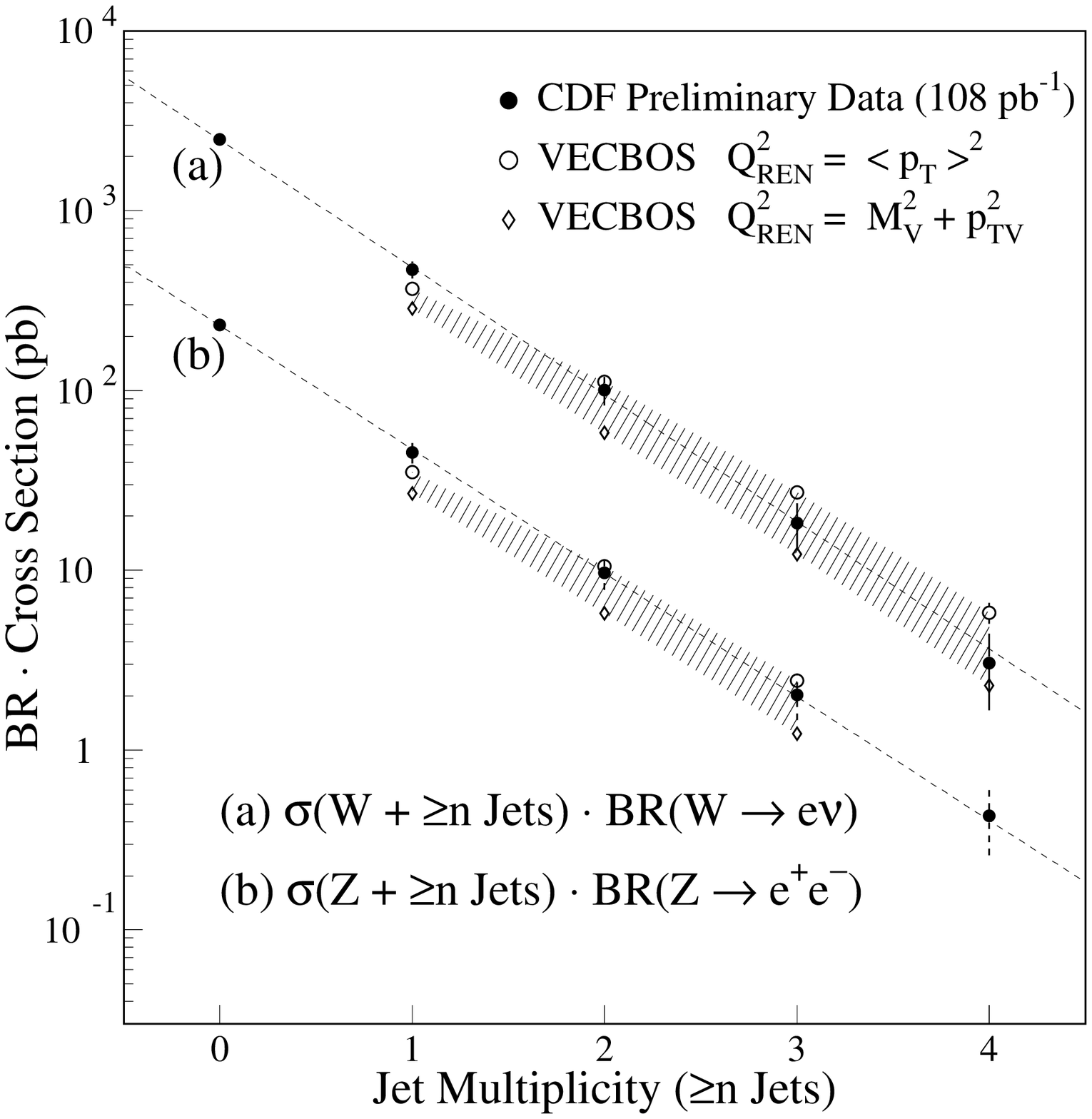,width=.45\textwidth}}\quad
        \subfigure[${\cal R}^{10}$ vs. $E_T^{min}$]
{\psfig{figure=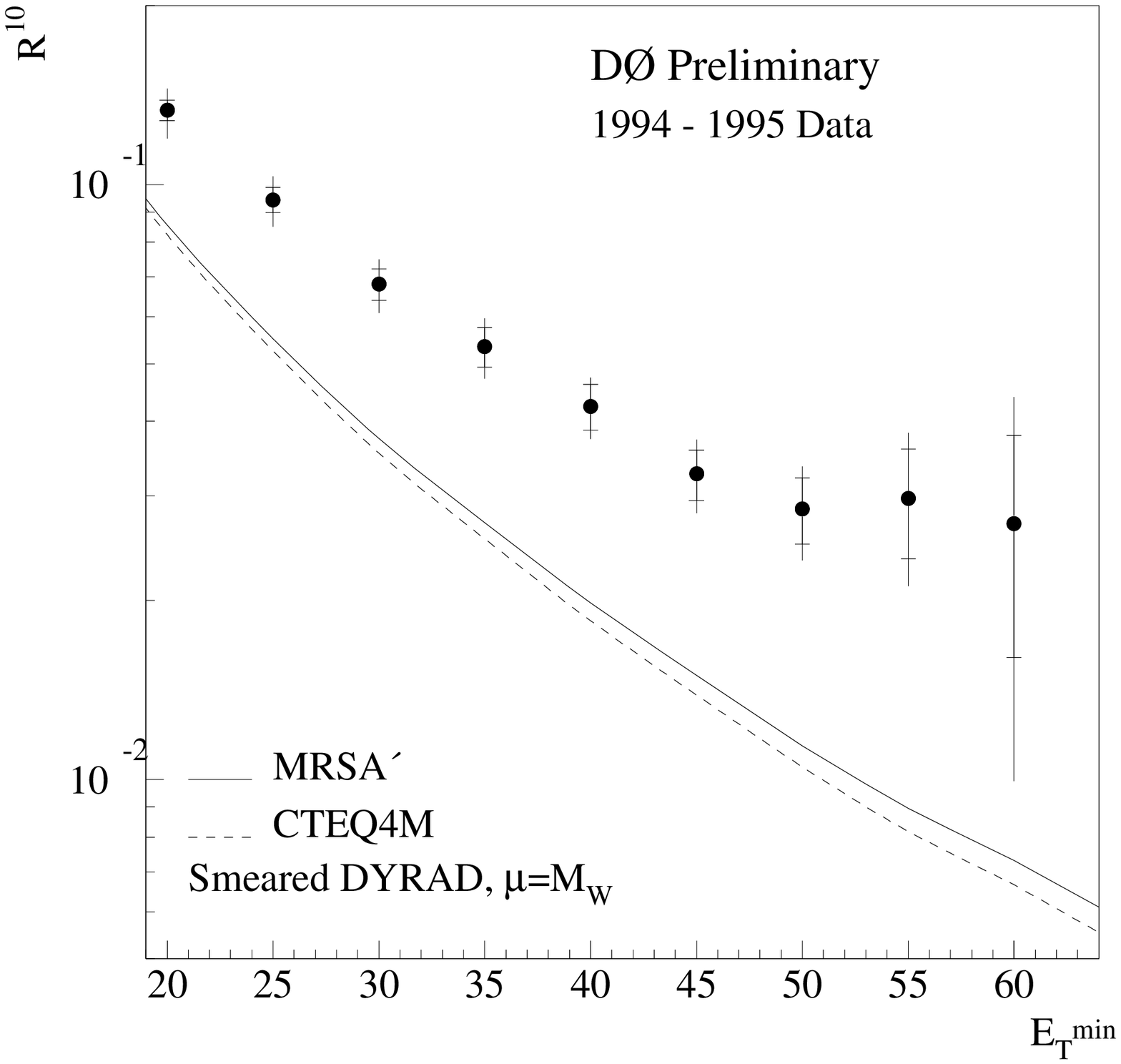,width=.45\textwidth}} 
}
  \caption{CDF and D\O\ results for $W/Z + $ Jets production.}
  \label{fig:wzjets}
\end{figure}

    The first analysis from the CDF collaboration was a measurement of the 
$W/Z + \geq n$ Jets cross sections for $n=1-4$.  Figure \ref{fig:wzjets}(a)
shows the inclusive associated jet multiplicity distribution for $W$ and $Z$
bosons.  The uncorrected data are compared to the CDF detector simulation
incorporating the {\small {VECBOS}}~\cite{vecbos} 
LO QCD calculation plus {\small {HERWIG}}~\cite{herwig59} 
parton shower and hadronisation. 
The {\small {CTEQ3M}}~\cite{cteq3m} 
parton densities were used.  The band in the theoretical
predictions represents the effect of varying the renormalisation and 
factorisation scales from $Q^2=M^2+p_T^2$ of the boson to the 
$\langle p_T\rangle^2$ of the partons.  Using the hard scale, $M^2+p_T^2$, the
LO QCD predictions are about a factor 1.7 lower than the data, for all jet 
multiplicities.  On the other hand the predictions for the softer scale,
$\langle p_T\rangle^2$, are in better agreement with the data.

    The D\O\ collaboration reported results on the ratio of the production
cross sections for $W + 1$ Jet to $W + 0$ Jets, ${\cal R}^{10}$, as a function
of the minimum jet transverse energy, shown in Fig. \ref{fig:wzjets}(b).  The
data between 20 and 60 GeV are consistently higher than the DYRAD NLO 
predictions by about a factor of two.  This 
is a rather curious result since it is in a domain where one generally expects 
QCD to work well.  

    D\O\ also investigated color coherence effects in $W + $ Jets events.  
For this study events with a $W$ boson
and opposing jet were selected and the distribution of soft particles
around the colorless $W$ boson and the jet (colored parton) 
was measured.  The color coherence signal is observed by comparing the 
multiplicity distributions of calorimeter towers with $E_T > 250$ MeV 
around the $W$ and around the jet.
It is concluded that both angular ordering and 
string fragmentation are needed 
in PYTHIA~\cite{jetset74} to describe the data.



\newpage
\section*{Conclusions}
The development of NLO calculations for a wide range of hadronic final state
variables and the increasingly precise data from HERA, the Tevatron and LEP
has provided a detailed testing ground for the strong interaction.
At the DIS97 workshop, beautiful agreement of the data from LEPII with QCD
was presented. At HERA and the Tevatron various chinks in the armour of QCD 
were identified. 
The detailed comparison of these data with the latest developments in the 
theoretical framework will determine whether
the established paradigms are sufficient to understand the many facets of
hadronic final state production reported at this workshop.

\section*{Acknowledgements}
It is a pleasure to thank all the speakers who contributed to the various
sessions, as well as those who contributed to the lively discussions following
these talks. The success of this working group was in large part due to the
outstanding organisation of the workshop by J. Repond and his team.

\end{document}